\documentclass[superscriptaddress,nofootinbib,twocolumn,aps,pra,longbibliography]{revtex4-2}
\usepackage[normalem]{ulem}
\usepackage{epsfig,amsmath,amssymb,color,algorithm,algcompatible,amsthm,bm}
\usepackage{multirow}
\usepackage{mathbbol}
\usepackage{comment}
\usepackage[english]{babel}
\usepackage[whole,autotilde]{bxcjkjatype} %
\usepackage{booktabs}
\usepackage{nicematrix}
\usepackage{animate}
\usepackage{tabularx}
\usepackage[colorinlistoftodos,prependcaption]{todonotes}

\newcommand{\bra}[1]{\langle{#1}|}
\newcommand{\ket}[1]{|{#1}\rangle}

\newcommand{\fig}[4]{
\begin{figure*}[ht]
\centering
\includegraphics[#4]{#1}
\caption{#2}
\label{#3}
\end{figure*}
}

\newtheorem{th.}{Theorem}
\newtheorem{co.}{Corollary}

\begin{document}

\title{
	Quantum algorithm for Electromagnetic Field Analysis
}

\author{Hiroyuki~Tezuka}
\email[]{hiroyuki.tezuka@keio.jp}
\affiliation{Advanced Research Laboratory, Sony Group Corporation, 1-7-1 Konan, Minato-ku, Tokyo, 108-0075, Japan}
\affiliation{Quantum Computing Center, Keio University, 3-14-1 Hiyoshi, Kohoku-ku, Yokohama, Kanagawa, 223-8522, Japan}

\author{Yuki Sato}
\email[]{yuki-sato@mosk.tytlabs.co.jp}
\affiliation{Toyota Central R\&D Labs., Inc., 1-4-14, Koraku, Bunkyo-ku, Tokyo, 112-0004, Japan}
\affiliation{Quantum Computing Center, Keio University, 3-14-1 Hiyoshi, Kohoku-ku, Yokohama, Kanagawa, 223-8522, Japan}

\begin{abstract}

Partial differential equations (PDEs) are central to computational electromagnetics (CEM) and photonic design, but classical solvers face high costs for large or complex structures. Quantum Hamiltonian simulation provides a framework to encode PDEs into unitary time evolution and has potential for scalable electromagnetic analysis.
We formulate Maxwell's equations in the potential representation and embed governing equations, boundary conditions, and observables consistently into Hamiltonian form. A key bottleneck is the exponential growth of Hamiltonian terms for complex geometries; we examine this issue and show that logical compression can substantially mitigate it, especially for periodic or symmetric structures. As a proof of concept, we simulate optical wave propagation through a metalens and illustrate that the method can capture wavefront shaping and focusing behavior, suggesting its applicability to design optimization tasks.
This work highlights the feasibility of Hamiltonian-based quantum simulation for photonic systems and identifies structural conditions favorable for efficient execution.

\end{abstract}

\maketitle

\section{Introduction}
\label{SEC:intro}

Computer-aided engineering (CAE) is a fundamental tool in modern product development, improving design efficiency, reducing cost, and enhancing product reliability.  
With advances in communication and sensing technologies, CAE, specifically electromagnetic field analysis, has become essential for the design of antennas, filters, waveguides, and integrated photonics. 
It also underpins electromagnetic compatibility for automotive and aerospace systems, radio-frequency safety assessment for medical devices, and the design of optical devices and metamaterials~\cite{taflove2005computational,jin2015finite,bowers2016recent,Jeong:24}.
Given these broad applications, improving the accuracy and scalability of electromagnetic CAE is a matter of significant industrial importance.

High-accuracy electromagnetic analysis, however, often requires substantial computational resources. 
In standard schemes such as the finite-difference time-domain (FDTD) method and the finite element method (FEM), refining the spatial resolution rapidly increases the number of grid points, while stability criteria force the time step to shrink proportionally; together, these effects cause the overall computational burden to grow explosively. 
Consequently, long-time simulations with large memory requirements are needed, beyond the capabilities of conventional computing resources.~\cite{weinan2007heterogeneous,6355594}.
Similar challenges arise across other CAE domains, e.g., structural vibration analysis for bridges and aircraft~\cite{mukhopadhyay2021structural} and the study of turbulence and unsteady flows~\cite{ferziger2002computational,pope2001turbulent,wilcox1998turbulence}, all of which require formidable computational costs.

Quantum computing has recently emerged as a promising framework for CAE, specifically for solving partial differential equations (PDEs). 
Several fundamental works proposed quantum computing-based approaches for solving PDEs based on quantum algorithms such as quantum linear system solvers and Hamiltonian simulation.
In particular, Hamiltonian simulation has recently attracted much attention owing to its efficient computation of time evolution by mapping physical laws to a quantum-mechanical form.
Although these fundamental works showed the potential of quantum computing for accelerating CAE, more practical applications should be studied to develop workflows and subroutines suitable for industrial deployment.

In this work, we demonstrate an application of quantum algorithms to electromagnetic field analysis, focusing on optical systems such as metalenses.
Maxwell's equations are mapped into the Hamiltonian simulation framework, enabling us to model wave propagation and focal-spot formation within a quantum circuit. We provide a proof-of-concept showing how quantum simulation can be applied to device-level optical design.
A critical consideration for practical applications is the rapid growth in the number of Hamiltonian terms when discretizing complex geometries. To investigate this issue, we evaluated typical structural patterns and confirmed that logical compression can effectively mitigate the number of Hamiltonian terms. Notably, the compression efficiency is especially strong for periodic and symmetric structures, which are typical in real-world optical components such as metasurfaces and photonic devices.
Through numerical experiments, we reproduce wave propagation and focusing consistent with classical simulations, while also clarifying how Hamiltonian compression interacts with device geometry. These results suggest that Hamiltonian simulation, combined with structural compression techniques, offers a promising basis for future quantum-assisted design workflows in computational electromagnetics.

The remainder of this paper is organized as follows. We first present the formulation of Maxwell's equations in a Hamiltonian framework, and describe the mapping to qubit registers.
Then, we demonstrate numerical experiments on a toy modeled optical system. Finally, we summarize the implications for optical device design and outline directions for future research.

\section{Mathematical Formulation}
\label{SEC:math}

In this section, we derive the governing equations used in electromagnetic field analysis, i.e., Maxwell’s equations, in a form suitable for quantum algorithms. 
As prior work, methods based on Schrödingerization have been proposed to directly analyze electromagnetic fields~\cite{refId0,ma2024schr}. 
In contrast, the present study adopts a potential-based formulation.

\subsection{Potential-based representation of Maxwell's equations}

Maxwell's equations, which govern the electromagnetic field in a homogeneous, isotropic, and lossless medium, can be written as follows:
\begin{align}
\nabla \cdot \bm{E} &= \frac{\rho}{\epsilon}, \quad 
\nabla \cdot \bm{B} = 0, \\
\nabla \times \bm{E} &= - \frac{\partial \bm{B}}{\partial t}, \quad 
\nabla \times \bm{B} = \mu \bm{J} + \mu \epsilon \frac{\partial \bm{E}}{\partial t},
\end{align}
where $\bm{E}$ and $\bm{B}$ denote the electric field vector and the magnetic flux density, respectively; $\epsilon$ and $\mu$ denote the permittivity and permeability of a medium; $\rho$ and  $\bm{J}$ denote the charge and current density vectors, respectively; and $\nabla=(\frac{\partial}{\partial x}, \frac{\partial}{\partial y}, \frac{\partial}{\partial z} )^\top$.
In this formulation, the electric field $\bm{E}$ and magnetic flux density $\bm{B}$ must be directly discretized and solved numerically. However, it is well known that numerical errors often violate the constraint $\nabla \cdot \bm{B}=0$ (the divergence-free condition of the magnetic field), which can lead to physically inconsistent results~\cite{MUNZ2000484,avazpour2025field}. 
This issue has also been pointed out in recent studies employing Schrödingerization for solving Maxwell's equations as a Hamiltonian system: in addition to preserving energy conservation through unitarity, maintaining $\nabla \cdot \bm{B}=0$ remains a central challenge in numerical simulations~\cite{refId0}. 
Thus, the treatment of the divergence constraint is an inherent issue in numerical electromagnetic field analysis.

To overcome this difficulty, we introduce the vector potential $\bm{A}$ and scalar potential $\phi$, defined such that:
\begin{align}
\bm{B} &= \nabla \times \bm{A} \label{eq:B}, \\
\bm{E} &= - \nabla \phi - \frac{\partial \bm{A}}{\partial t}. \label{eq:E}
\end{align}
This guarantees that $\nabla \cdot \bm{B}=0$ is automatically satisfied, eliminating the need for numerical correction of the divergence condition. Furthermore, to analyze dynamical electromagnetic fields, we impose the Lorenz gauge condition
\begin{align}
\nabla \cdot \bm{A} + \mu \epsilon \frac{\partial \phi}{\partial t} = 0, \label{eq:guage}
\end{align}
under which Maxwell's equations can be expressed in the form of wave equations.
Specifically, the electromagnetic potentials satisfy the following equations:
\begin{align}
\left(\nabla^2 -\frac{1}{c^2} \frac{\partial^2}{\partial t^2} \right) \bm{A} &= -\mu \bm{J} \label{eq:wave_A} \\
\left(\nabla^2 -\frac{1}{c^2} \frac{\partial^2}{\partial t^2} \right) \phi &= -\frac{\rho}{\epsilon}, \label{eq:wave_phi}
\end{align}
where the propagation speed of the electromagnetic wave in the medium is given by $c = 1 / \sqrt{\mu \epsilon}$.
This formulation separates the variables $\bm{A}$ and $\phi$.
That is, once the charge distribution $\rho$ and current distribution $\bm{J}$ are given, $\bm{A}$ and $\phi$ can be obtained independently by wave equations, and the physical electric and magnetic fields can then be consistently reconstructed from these potentials.
It should be noted that whether Eq.~\eqref{eq:wave_A} or Eq.~\eqref{eq:wave_phi} is solved depends on the specific problem of interest. 
The two are complementary and share the same mathematical structure; however, when treating dynamical electromagnetic fields, it is preferable to focus on the vector potential as the primary variable and solve Eq.~\eqref{eq:wave_A}. 
This is because Eq.~\eqref{eq:wave_phi} accounts only for the electric field component derived from charges, and does not capture the contributions from magnetic fields nor guarantee the divergence-free condition of the magnetic field.
By contrast, solving Eq.~\eqref{eq:wave_A} and reconstructing the electromagnetic fields via Eq.~\eqref{eq:E}, together with the gauge condition Eq.~\eqref{eq:guage}, naturally yields the required electromagnetic quantities.

\subsection{Mapping onto Hamiltonian}

In this study, we adopt Eq.\eqref{eq:wave_A} as the governing equation of the system and embed it into a Hamiltonian framework, following the methods proposed in previous works~\cite{PhysRevResearch.6.033246,PhysRevApplied.23.014063}, in order to simulate the time evolution of the system. 
For simplicity, let us first consider the case without any external current (i.e., $\bm{J} = \bm{0}$). 
Under this assumption, Eq.~\eqref{eq:wave_A} can be transformed into the following set of first-order differential equations by means of variable substitution:
\begin{align}
\frac{\partial \bm{\psi}(t, \bm{x})}{\partial t} = -i\mathcal{H} \bm{\psi}(t, \bm{x}) \label{eq:time_evolution},
\end{align}
where
\begin{align}
\bm{\psi}(t, \bm{x}) = \begin{pmatrix} \frac{1}{c} \dot{\bm{A}} \\ \nabla \bm{A} \end{pmatrix}, \label{eq:vec_psi} \\
\mathcal{H} = \begin{pmatrix} 0 & -c \nabla^\top \\
                              -c \nabla & 0  \end{pmatrix}. \label{eq:mat_ham}                            
\end{align}
We denote $\nabla \bm{A} =  (\nabla A_x, \nabla A_y, \nabla A_z)^\top = (\partial_x A_x, \partial_y A_x, \partial_z A_x, \partial_x A_y, \partial_y A_y, \partial_z A_y, \partial_x A_z, \partial_y A_z, \partial_z A_z)^\top$, and note that Eq.~\eqref{eq:vec_psi} includes 12-dimension vector.

Subsequently, these equations are discretized and expressed using qubits. 
As a preliminary step, we consider the discretization of a scalar field $u(\bm{x})$ using $N=2^n=2^{n_x+n_y+n_z}$ grid points. 
The scalar field $u(\bm{x})$ is then represented at each spatial node as $\bm{u}:=[u(\bm{x}^{[0]}), u(\bm{x}^{[1]}), \dots, u(\bm{x}^{[N-1]})]^\top$, where $\bm{x}^{[j]}$ denotes the $j$-th node. The discretized scalar field $\bm{u}$ is mapped onto the following state vector:
\begin{align} \label{eq:ket_u}
    \ket{u} &:= \frac{1}{\| \bm{u} \|} \sum_{j=0}^{2^n-1} u(\bm{x}^{[j]}) \ket{j} \\
            &= \frac{1}{\| \bm{u} \|} \sum_{j_x=0}^{2^{n_x}-1} \sum_{j_y=0}^{2^{n_y}-1} \sum_{j_z=0}^{2^{n_z}-1} u(\bm{x}^{[j_x, j_y, j_z]}) \ket{j_x} \ket{j_y} \ket{j_z},
\end{align}
where $\ket{j}:=\ket{j_{n-1} j_{n-2} \dots j_0}$ with $j_{n-1}, j_{n-2}, \dots, j_0 \in \{0, 1\}$ represents the computational basis state, and $N_\mu=2^{n_\mu}$~($\mu \in \{x,y,z\}$) denotes the number of subdivisions along the $x_\mu$-axis.

Using this representation, the finite-difference expression for the spatial derivative $\partial/\partial x_\mu$ can be derived as follows. 
For example, in a three-dimensional orthogonal coordinate system with unit vectors along the $x_\mu$-axis denoted by $\bm{e}_\mu$ and grid spacing $h$, the forward difference of a function $u(\bm{x})$ in the $x_\mu$-direction can be written as
\begin{align}
\frac{\partial u(\bm{x}^{[j]})}{\partial x_\mu } \approx \frac{u_{\mu,j+1} - u_j}{h}.
\end{align}
These operators can, in turn, be represented in terms of the shift operator $S_\mu^{-}$, which acts only on the qubits encoding the $x_\mu$-coordinate, and the identity operator $I$, as follows (for the case of Dirichlet boundary conditions):
\begin{align}
    \frac{1}{h} \left( S_\mu^{-} - I^{\otimes n} \right) \sum_{j=0}^{2^n-1} u(\bm{x}^{[j]}) \ket{j} = \sum_{j=0}^{2^n-1} \frac{ u_{\mu, j+1} - u_j }{ h } \ket{j},
\end{align}
where $u(\bm{x}^{[j]} + h\bm{e}_\mu) = u_{\mu, j+1}, u(\bm{x}^{[j]}) = u_j$.
As shown in prior work~\cite{PhysRevResearch.6.033246}, backward and central difference operators can be expressed in a similar manner.

With these definitions, the extension to a vector field $\bm{A}(\bm{x})$ is straightforward, and Eq.~\eqref{eq:time_evolution}, \eqref{eq:vec_psi}, and \eqref{eq:mat_ham} can be explicitly expressed in terms of qubits. 
Furthermore, in order to represent the twelve components of Eq.~\eqref{eq:vec_psi} in a compact form, we introduce an index state
$4 = \lfloor\log_2 12\rfloor$ qubits, yielding the following representation:
\begin{align}
    \frac{\partial \ket{\psi(t)}}{\partial t} = -i\mathcal{H} \ket{\psi(t)}, \label{eq:time_evolution_ket}
\end{align}
where
\begin{widetext}
\begin{align}
&\ket{\psi(t)} = \sum_{\mu'=0}^{11} \sum_{j_x=0}^{N_x-1} \sum_{j_y=0}^{N_y-1} \sum_{j_z=0}^{N_z-1} u_{\mu'}(t, \bm{x}^{[j_x, j_y, j_z]}) \ket{\mu'} \ket{j_x} \ket{j_y} \ket{j_z}, \label{eq:ket_psi} \\
&\mathcal{H}
= i \begin{bNiceMatrix}[columns-width=auto]
\Block{3-3}{\bm{0}} &  &  & \tilde{c}(\bm{x}) D^+_x & \tilde{c}(\bm{x}) D^+_y & \tilde{c}(\bm{x}) D^+_z & 0 & 0 & 0 & 0 & 0 & 0 \\
 &  &  & 0 & 0 & 0 & \tilde{c}(\bm{x}) D^+_x & \tilde{c}(\bm{x}) D^+_y & \tilde{c}(\bm{x}) D^+_z & 0 & 0 & 0 \\
 &  &  & 0 & 0 & 0 & 0 & 0 & 0 & \tilde{c}(\bm{x}) D^+_x & \tilde{c}(\bm{x}) D^+_y & \tilde{c}(\bm{x}) D^+_z \\
D^-_x \tilde{c}(\bm{x}) & 0 & 0 & \Block{8-8}{\bm{0}} \\
D^-_y \tilde{c}(\bm{x}) & 0 & 0 &  &  &  &  &  &  &  &  &  \\
D^-_z \tilde{c}(\bm{x}) & 0 & 0 &  &  &  &  &  &  &  &  &  \\
0 & D^-_x \tilde{c}(\bm{x}) & 0 &  &  &  &  &  &  &  &  &  \\
0 & D^-_y \tilde{c}(\bm{x}) & 0 &  &  &  &  &  &  &  &  &  \\
0 & D^-_z \tilde{c}(\bm{x}) & 0 &  &  &  &  &  &  &  &  &  \\ 
0 & 0 & D^-_x \tilde{c}(\bm{x}) &  &  &  &  &  &  &  &  &  \\
0 & 0 & D^-_y \tilde{c}(\bm{x}) &  &  &  &  &  &  &  &  &  \\
0 & 0 & D^-_z \tilde{c}(\bm{x}) &  &  &  &  &  &  &  &  &  \end{bNiceMatrix}. \label{eq:ketbra_ham}
\end{align}
\end{widetext}
Here, the finite-difference operators $D^{-}_{\mu}$ and $D^{+}_{\mu}$ are defined together with the position-dependent material constant $\tilde{c}(\bm{x})$ as
\begin{align}
D^{-}_{\mu} &= \frac{1}{h} \left( S_{\mu}^{-} - I^{\otimes n} \right), D^{+}_{\mu} = \frac{1}{h} \left(I^{\otimes n} - S_{\mu}^{+} \right), \\
\tilde{c}(\bm{x}) &= \sum_{j_x=0}^{N_x-1} \sum_{j_y=0}^{N_y-1} \sum_{j_z=0}^{N_z-1} c(\bm{x}^{[j_x, j_y, j_z]}) \ket{j_x, j_y, j_z} \bra{j_x, j_y, j_z}.
\end{align}

From the above formulation, the number of qubits required for a lattice of size 
$N$ scales as $O(\log_2N)$. 
For example, in the case of a three-dimensional grid of size approximately $10^3 \times 10^3 \times 10^3$, the required number of qubits is 34. 
This indicates that, compared with classical computation, the proposed approach can achieve exponential improvement in terms of scaling with respect to the number of grid points.

\subsection{Boundary condition} %

Boundary conditions can also be expressed within the Hamiltonian.
Specifically, additional terms corresponding to the boundary conditions are incorporated into the finite-difference operators. 
For instance, in the case of forward differences, the Neumann boundary condition and the periodic boundary condition can be represented respectively as follows:
\begin{align}
D_N^+ &:= \frac{1}{h} (S^- -I^{\otimes n} + \sigma_{11}^{\otimes n}) \\
D_P^+ &:= \frac{1}{h} (S^- -I^{\otimes n} + \sigma_{10}^{\otimes n})
\end{align}
Similar modifications apply to the other finite-difference operators as well. Further details can be found in~\cite{PhysRevResearch.6.033246,PhysRevApplied.23.014063}.

\subsection{Observable}
Although Hamiltonian-simulation-based approaches allow analyses with exponentially high resolution in terms of the number of qubits, reconstructing the full information over the entire spatial domain requires quantum state tomography~\cite{PhysRevA.65.012301}, which incurs exponential overhead. 
Hence, we need to carefully select some observables to extract meaningful information from the quantum state after time evolution.
In practice, it is sufficient to extract a quantity such as figures of merit (FoMs) or other derived metrics evaluated over specific regions of interest, rather than recovering the complete high-resolution information.
Thus, we consider a physical quantity represented by an observable $O$, and define $\mathcal{X}$ as the index set of grid points on which we evaluate the integrated value of $O$.
That is, we focus on the following quantity $O_\mathrm{\mathcal{X}}$:
\begin{align}
O_{\mathcal{X}} = \sum_{\bm{x} \in \mathcal{X}} O(\bm{x}^{[j_x, j_y, j_z]}).
\end{align}
Conceptually, this observable corresponds to counting the measurement outcomes of the spatial register that fall within the specified region of interest represented by $\mathcal{X}$. 
The larger the integrated region is, the more qubits unrelated to specifying the region can be traced out, thereby improving sampling efficiency.

Since spatial integration is generally performed over continuous domains, logical compression techniques, originally developed for Hamiltonian compression~\cite{PhysRevApplied.23.014063}, can also be applied to compress the observables. 
Although we do not pursue specific applications here, identifying the types of spatial structures that admit efficient compression and are well-suited for quantum algorithms remains an important topic for further research.

\section{Conceptual Demonstration: Metalens Design} 
\label{SEC_poc}
In this section, as a proof of concept, we address an optical simulation problem based on the theoretical framework described above.

\subsection{Background and motivation}
Optical technologies play a crucial role in modern society, serving a wide range of applications such as information and communication systems, imaging, sensing, and display devices.
Owing to the growing demand for miniaturization, integration, and energy efficiency, intensive research has focused on emerging photonic platforms that overcome the limitations of conventional refractive and diffractive optics. 
In this context, metasurfaces and metalenses, which are ultrathin, planar optical components with subwavelength-scale nanostructures, have emerged as promising candidates to replace conventional bulky traditional components~\cite{bowers2016recent}. 
Their capability to manipulate light in amplitude, phase, and polarization with high precision provides new opportunities for imaging, consumer electronics, and quantum optical devices~\cite{ali2022metamaterials,Jeong:24}.
Nevertheless, their design remains a highly challenging problem. The behavior of each nanounit (meta-atom) strongly depends on coupling with its neighbors, and the overall optical response arises from collective interactions across multiple scales~\cite{lee2024data,li2022empowering}.

In this study, we focus on the challenge of simulation resources. 
Accurately describing the propagation of optical waves through a metalens requires a consistent treatment from the microscopic local response to macroscopic wavefront formation, a task that demands enormous computational resources on classical platforms. 
For this type of multiscale computational burden, quantum computing---particularly Hamiltonian simulation---offers a promising new avenue. 
Building on this background, we attempt a proof of concept in which Hamiltonian simulation is used to reproduce wave propagation through a metalens and to observe the focusing process, thereby demonstrating a new design framework enabled by quantum algorithms.

\subsection{Numerical experiment}
In this section, we formulate metalens design as a focal position optimization problem. Specifically, we present a toy model in which the design parameters of a lens are optimized such that the incident wave energy is concentrated at a designated focal position. 
Although realizing a broad range of optical functionalities ideally involves optimizing multiple phenomena, such as interference, dispersion, and refraction, we restrict our attention to the refractive-index distribution, treating it as a toy problem with minimal degrees of freedom for simplicity.

First, we simplify the governing equations. In particular, in Eq.~\eqref{eq:time_evolution_ket}, \eqref{eq:ket_psi}, and \eqref{eq:ketbra_ham}, we restrict our analysis to a two-dimensional spatial domain while assuming that the electromagnetic field is uniform along the remaining one dimension. Physically, this corresponds to the TM mode, and combined with the assumption of no charge distribution ($\rho = 0$), we obtain the relation $\bm{E}= E_z\bm{e}_z = -\partial A_z/\partial t~ \bm{e}_z$.
Under this setting, the objective is to estimate the configuration that maximizes the electromagnetic energy within the monitoring region $\mathcal{X}$. The governing equation to be solved in this case is given by Eq.~\eqref{eq:time_evolution_ket}, which can be expressed as follows:
\begin{align}
\ket{\psi(t)} &= \sum_{\mu'=0}^{2} \sum_{j_x=0}^{N_x-1} \sum_{j_y=0}^{N_y-1} u_{\mu'}(t, \bm{x}^{[j_x, j_y]}) \ket{\mu'} \ket{j_x} \ket{j_y}, \label{eq:ket_psi_exp} \\
\mathcal{H}
&= i \begin{bNiceMatrix}[columns-width=auto]
0 & D^+_x c(\bm{x}) & D^+_y c(\bm{x}) \\
c(\bm{x}) D^-_x & 0 & 0 \\
c(\bm{x}) D^-_y & 0 & 0 \end{bNiceMatrix}, \label{eq:ketbra_ham_exp}
\end{align}
It should be noted here that $u_{\mu'=0}$ corresponds to $E_z / c(\bm{x})$.

The optimization target in this formulation is the spatial distribution of the propagation velocity $c(\bm{x})$.
Ideally, one would design a continuous refractive index profile; however, in practice it is difficult to fabricate materials with arbitrary optical properties.
In this study, to approximate a continuous refractive index distribution, we adopt a binary discretization scheme: each spatial point is assigned either free-space velocity $c=1.0$ or high-index TiO$_2$ velocity $c=0.45$. 
Effective intermediate refractive indices are reproduced by smoothing the local average over a $3\times3$ pixel region. 
This binary approximation corresponds to the common effective medium approach in metasurface design, where subwavelength patterning emulates graded index profiles~\cite{hassan2020integrated,PhysRevE.81.046607,Panipinto:24}.
The spatial domain is discretized on a $64 \times 64$ square lattice. For the purpose of evaluating the efficiency of logical compression in the following section, a $128 \times 128$ grid is also considered.

\begin{figure}[h!] 
   \includegraphics[width=0.35\textwidth]{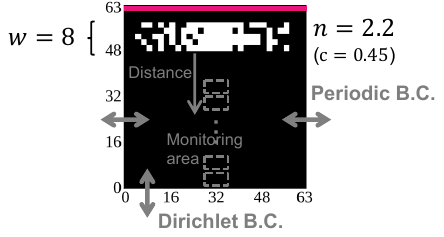}
   \caption{Settings}
   \label{fig:settings}
\end{figure}

The input wave is modeled as an impulsive plane wave incident from the upper boundary (red line in Fig.~\ref{fig:settings}). The electromagnetic energy intensity at each coordinate resulting from this input is used as the measurement target. Accordingly, the observable $O_\mathcal{X}$ corresponding to the monitoring region $\mathcal{X}$ is defined as follows:
\begin{align}
O_\mathcal{X} &= \sum_{(j_x, j_y) \in \mathcal{X}} \ket{0}\ket{j_x}\ket{j_y}\bra{0}\bra{j_x}\bra{j_y}. 
\end{align}
More specifically, in the top monitoring area (the dashed rectangle in Fig.~\ref{fig:settings}),
\begin{align}
\mathcal{X} &= \{(0111\mathrm{xx}, 1010\mathrm{xx})_2, (1000\mathrm{xx}, 1010\mathrm{xx})_2 \}
\end{align}
corresponds to an integration over 32 pixels. For evaluating other focal positions, the monitoring region corresponding to 32 pixels is shifted accordingly, and the values are accumulated. Here, the coordinates in the equation are expressed in binary notation, and $\mathrm{x}$ denotes a ``don't care'' bit (0 or 1).

In embedding the dynamics into the quantum circuit, we employed the first-order Trotter decomposition~\cite{trotter1959product,suzuki1976generalized}. The temporal discretization width in the simulations was set to $\Delta t=0.01$ for both the classical FDM method and the Hamiltonian simulation. All quantum simulations were implemented using Qiskit 1.0~\cite{qiskit2024}.

\subsection{Pre-analysis: Compression efficiency of Hamiltonian terms}
\fig{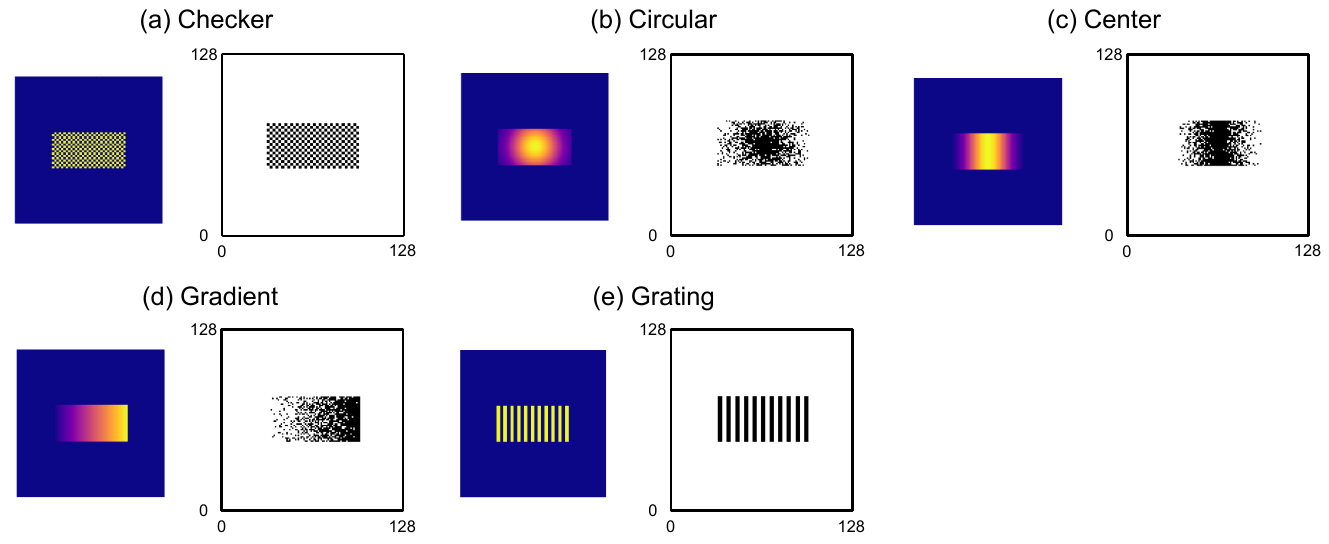}
{Binary-approximated structures. Left: original refractive index distribution (maximum value shown in yellow). Right: binary-approximated distribution.}
{fig:compression}{width=0.90\textwidth}

The computational complexity strongly depends on the number of terms in the Hamiltonian. Previous studies~\cite{PhysRevApplied.23.014063} have shown that, although heuristic in nature, the method of logical compression can drastically reduce the number of Hamiltonian terms for representing the spatial distribution of $c(\bm{x})$. In the present case, the number of terms is determined by the number of lattice sites occupied by the material, and thus the efficiency of compression is expected to depend significantly on the structural features of the optical components under consideration. To evaluate this effect, we first examined the performance of logical compression on several representative structures.

The results of applying logical compression to the five patterns illustrated in Fig.~\ref{fig:compression} are summarized in Table~\ref{tab:copmression}. 
As is evident from the table, structures exhibiting periodicity or geometric symmetry (i.e., checker and grating) allow for substantial reductions in the number of Hamiltonian terms. 
By contrast, in cases where the refractive index varies continuously, the compression efficiency deteriorates significantly. 
These findings indicate that periodic or geometrically symmetric structures are particularly well-suited to the proposed algorithm.

It is worth noting that in practical metasurface or metamaterial designs, periodic structures are often employed, making this result favorable from the perspective of real-world applicability. Moreover, even for structures that appear continuously varying at a macroscopic level, it may be possible to achieve significant improvements in compression efficiency by adopting representation schemes established in mature fields such as computational mechanics. Analogies from classical domains such as image compression may also provide valuable insights.

\begin{table}[h]
\centering
    \caption{The number of terms in Hamiltonian}
    \label{tab:copmression}
\begin{tabular}{|cc|c|c|c|c|c|}
\hline
\multicolumn{2}{|c|}{}                     &(a)&(b)&(c)&(d)&(e) \\ \hline
\multicolumn{1}{|c|}{\multirow{2}{*}{\# of terms}} & Before & 1024 & 974 & 995 & 1035 & 642   \\ \cline{2-7} 
\multicolumn{1}{|c|}{}                  & After & 8 & 463 & 483 & 483 & 44 \\ \hline
\multicolumn{2}{|c|}{Ratio}                     & 0.8\% & 48\% & 49\% & 47\% & 7\% \\ \hline
\end{tabular}
\end{table}

\subsection{Simulation of wave propagation}
Using Hamiltonian simulation, we reproduced the propagation of the wavefront through the lens. 
In this case, we employed a centrally focusing lens, which visualizes the focusing effect within a small computational domain, with thickness $w = 8$ (Fig.~\ref{fig:settings}). 
Figure~\ref{fig:result_qc} compares the computational results obtained with the proposed method and those from the classical finite-difference method (FDM). 
The results confirm both qualitative and quantitative agreement between the two approaches.

\fig{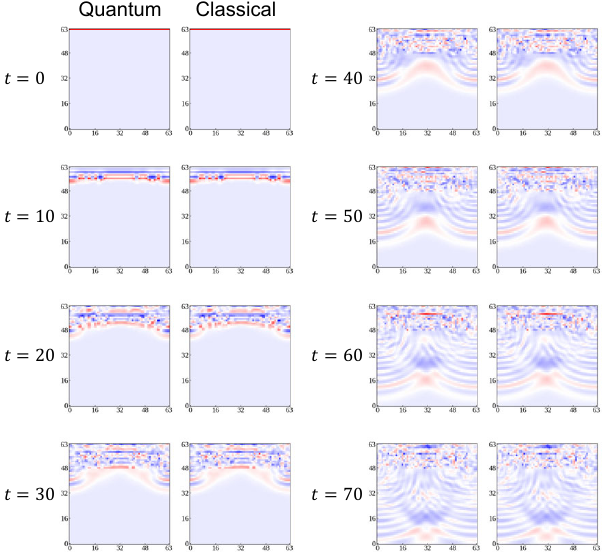}
{Wave propagation calculated by quantum and classical algorithm. In the figure, ``Quantum'' denotes Hamiltonian simulation and ``Classical'' denotes finite difference method. The color plots represents the amplitude of $E(\bm{x})$.}
{fig:result_qc}{width=0.65\textwidth}

\fig{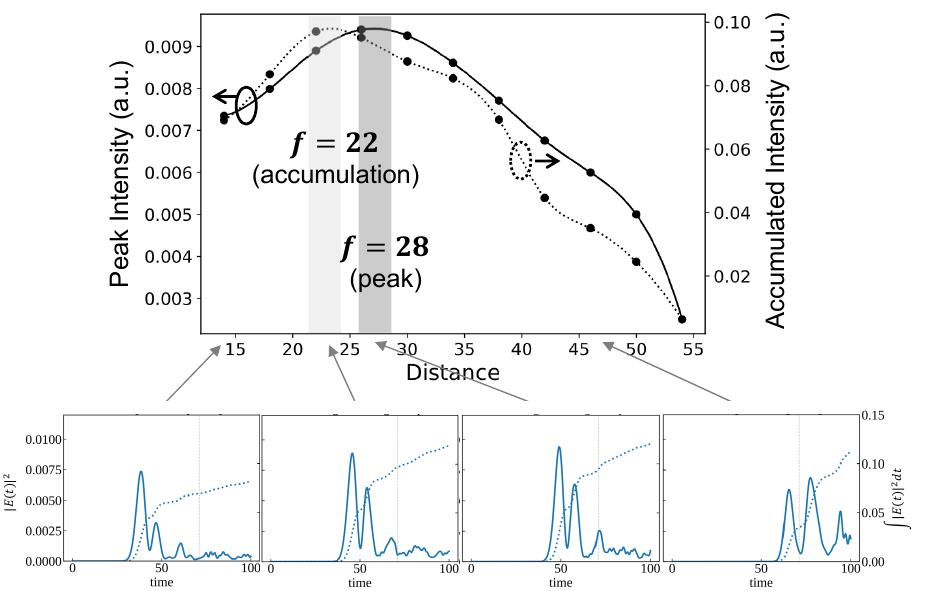}
{The location dependency of the power intensity $\int_{\mathcal{X}} |E(t, \bm{x})|^2d\bm{x}$ and the accumulated value $\int_t\int_{\mathcal{X}} |E(t, \bm{x})|^2d\bm{x}dt$. The integrated time window is from $t=0$ to $t=70$ to avoid the effect of reflection. The distance is the number of pixels from the bottom edge of the lens.} 
{fig:time_sweep}{width=0.5\textwidth}

Next, we identified the focal point. The focal position was evaluated by observing the intensity distribution at the monitoring points. For optical characterization, however, it is necessary to consider not only instantaneous peaks but also information derived from time-integrated values. Figure~\ref{fig:time_sweep} plots the power intensity at each position below the metasurface lens, based on the simulation results shown in Fig.~\ref{fig:result_qc}. The solid line represents the instantaneous peak intensity, while the dotted line corresponds to the time-integrated intensity. The former is particularly relevant for phenomena such as single-photon detection, whereas the latter is the quantity typically referenced in continuous-light measurements.

The subplots illustrate the temporal dependence of power at each monitoring position. These results reveal that the peak intensity exhibits a strong dependence on the observation location. This dependence arises from the spatially modulated overlap of equal-phase surfaces induced by the metasurface, which varies significantly with position. In other words, identifying the point at which the overlap of equal-phase surfaces is maximized, and where the energy is most strongly concentrated in space, is equivalent to determining the focal point.

From Fig.~\ref{fig:time_sweep}, it can be seen that the instantaneous peak occurs near $f=28$, while the integrated value peaks around $f=22$. In conclusion, for a metasurface lens of thickness $w=8$, the focal length is located in the range $f=22\sim 28$.
Here, the distance from the most strongly focused monitoring position to the edge of the metalens is denoted as $f$.
It should be noted that the analyses shown in Figs.~\ref{fig:result_qc}-\ref{fig:opt}, namely the spatiotemporal distributions of amplitudes obtained through time evolution by Hamiltonian simulation, are made feasible here because a state-vector simulator is employed. In order to perform such analyses on actual devices, it is necessary to efficiently extract the relevant information from the resulting quantum states. The development of concrete methods for this purpose constitutes an important direction for future research.

\fig{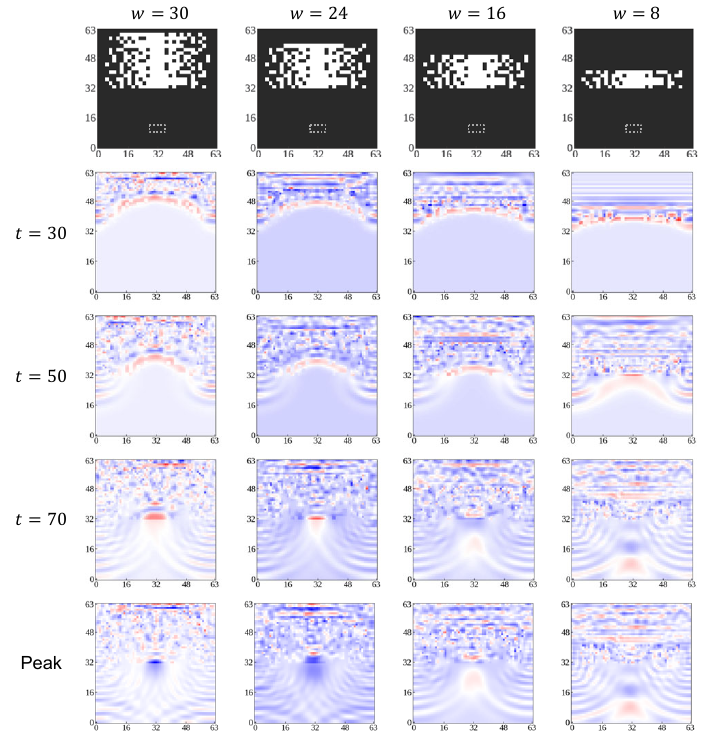}
{Wave propagation for metalenses with varying thicknesses. Colors are normalized in each plot, with the maximum value shown in red and the minimum in blue.}
{fig:opt}{width=0.65\textwidth}

\subsection{Focal point design}

Next, we investigated the effect of varying the lens thickness through numerical simulations. The results are shown in Fig.~\ref{fig:opt}. The top row illustrates the metasurface lens geometries considered, the middle row depicts the corresponding time evolutions, and the bottom row presents snapshots at the instants when the transmitted amplitude reaches its maximum. From these results, it is evident that the location of maximum energy concentration (i.e., the focal point) shifts closer to the lens as its thickness increases, consistently reflecting the expected physical behavior.

For reference, Table~\ref{tab:basic} summarizes the intensity values within the region enclosed by the dotted outline in the top-row figures (corresponding to distance $=24$). As observed in the previous section, the intensity near the focal distance for $w=8$ attains the highest value among the configurations considered.

\begin{table}[h]
\caption{Metric for power concentration (unit: a.u.) }
  \centering
  \begin{tabular}{|c|c|c|c|c|} %
  \hline
    Thickness $w$ & \parbox{1cm}{30} & \parbox{1cm}{24} & \parbox{1cm}{16} & \parbox{1cm}{8} \\ \hline
    Peak & 6.8 & 34 & 46 & 90 \\ \hline
    Accumulated  & 5.5 & 11 & 35 & 93 \\
  \hline
  \end{tabular}
  \label{tab:basic}
\end{table}

\subsection{Discussion}

When considering the application of the present study to metasurface lens design, the workflow would involve defining observables according to the design specifications, executing simulations under a variety of design parameters, and subsequently identifying those parameters that satisfy the specifications. The principal computational bottleneck in this process lies in performing time-domain and spatial sweeps, corresponding to the type of information illustrated in Fig.~\ref{fig:time_sweep}. Furthermore, if time integration is required, a large number of trials must be conducted with fine temporal discretization over the integration interval, followed by extensive sampling of the results. Developing algorithms capable of executing these procedures efficiently is thus of critical importance for the industrial application of quantum simulation, and represents a key direction for future research.

\section{Conclusion}
\label{SEC:conclusion}
We have presented a quantum algorithm for electromagnetic field analysis by formulating Maxwell's equations in a potential-based Hamiltonian framework and embedding them into quantum circuits, 
while boundary conditions and observables can also be expressed in Hamiltonian form. 
Numerical experiments on metalens systems demonstrated that Hamiltonian simulation can reproduce both wave propagation and focal behavior consistent with classical finite-difference results. 
Furthermore, logical compression proved effective in reducing Hamiltonian term counts, particularly for periodic or symmetric structures, highlighting the compatibility of discretized photonic designs with the proposed approach.
Although challenges remain in terms of information readout from the final state, such as designing effective observables and extracting the required information with limited sampling on real devices, the study demonstrates that quantum simulation offers a promising new paradigm for CAE.
In particular, its capability to directly capture multiscale electromagnetic interactions will serve as a powerful tool for future optical device and metasurface design. 
Building on the present proof of concept, future work will focus on integrating advanced decomposition techniques and exploring larger-scale implementations to enable practical quantum-assisted engineering workflows.

\section*{Acknowledgments}
The authors would like to thank Keio University Quantum Computing Center, especially Prof. Naoki Yamamoto, for fruitful discussions.

\bibliography{main}

\end{document}